\documentclass[twocolumn,english,pra,showpacs]{revtex4-1}
\usepackage[T1]{fontenc}
\usepackage[latin9]{inputenc}
\usepackage{amsmath}
\usepackage{graphicx}
\usepackage{amssymb}
\usepackage{esint}

\makeatletter
\@ifundefined{textcolor}{}
{%
 \definecolor{BLACK}{gray}{0}
 \definecolor{WHITE}{gray}{1}
 \definecolor{RED}{rgb}{1,0,0}
 \definecolor{GREEN}{rgb}{0,1,0}
 \definecolor{BLUE}{rgb}{0,0,1}
 \definecolor{CYAN}{cmyk}{1,0,0,0}
 \definecolor{MAGENTA}{cmyk}{0,1,0,0}
 \definecolor{YELLOW}{cmyk}{0,0,1,0}
 }

\makeatletter
\usepackage{dcolumn}
\usepackage{bm}
\usepackage{amsthm}\@ifundefined{definecolor}
 {\@ifundefined{definecolor}
 {\usepackage{color}}{}
}{}

\makeatother

\usepackage{babel}

\makeatother

\usepackage{babel}

\begin{document}
\newtheorem{conjecture}{Conjecture}\newtheorem{corollary}{Corollary}\newtheorem{theorem}{Theorem}
\newtheorem{lemma}{Lemma}
\newtheorem{observation}{Observation}
\newtheorem{definition}{Definition}\newtheorem{remark}{Remark}\global\long\global\long\def\ket#1{|#1 \rangle}
 \global\long\global\long\def\bra#1{\langle#1|}
 \global\long\global\long\def\proj#1{\ket{#1}\bra{#1}}

\title{Quantum measurements with preselection and postselection}

\author{Xuanmin Zhu{$^{1}$$^{,}$$^{2}$}}\email{zhuxuanmin@gmail.com} \author{Yuxiang Zhang{$^{2}$$^{,}$${^3}$}} \author{Shengshi Pang$^{2}$}\email{pangss@mail.ustc.edu.cn} \author{Chang Qiao$^{2}$} \author{Quanhui Liu$^{1}$}\email{quanhuiliu@gmail.com} \author{Shengjun Wu$^{2}$}\email{shengjun@ustc.edu.cn}

 \affiliation{$^{1}$School for Theoretical Physics, and Department of Applied Physics,
 Hunan University, Changsha 410082, China \\
 $^{2}$Hefei National Laboratory for Physical Sciences at Microscale
 and Department of Modern Physics, University of Science and Technology
 of China, Hefei, Anhui 230026, China   \\
 $^{3}$School for the Gifted Young, University of Science and Technology of China, Hefei, Anhui 230026, China}

\date{\today}

\pacs{03.65.Ta, 03.67.-a}
\begin{abstract}
We study quantum measurement with preselection and postselection, and derive the precise expressions of the measurement results without any restriction on the coupling strength between the system and the measuring device.  For a qubit system, we derive the maximum pointer shifts by choosing appropriate initial and finial states.  A significant amplification effect is obtained when the interaction between the system and the measuring device is very weak, and typical ideal quantum measurement results are obtained when the interaction is strong. The improvement of the signal-to-noise ratio (SNR) and the enhancement of the measurement sensitivity (MS) by weak measurements are studied. Without considering the probability decrease due to postselection, the SNR and the MS can be both significantly improved by weak measurements; however, neither SNR nor MS can be effectively improved when the probability decrease is considered.
\end{abstract}
\maketitle
\section{Introduction}
In 1988, Aharonov, Albert, and Vaidman introduced the idea of weak measurement~\cite{aav}.
For a very weak interaction between a system and measuring apparatus, they showed that the measurement results obtained can be much larger than the eigenvalues of the observable by selecting appropriate initial and final states. The counterintuitive results have been implemented in experiments~\cite{exp1,exp2}. Weak measurement has provided a new perspective to the famous Hardy's paradox~\cite{hardy1,hardy2}, and the illustrations have been realized in experiments~\cite{exp3}. Using the idea of weak measurement, Hosten and Kwiat have observed the tiny spin Hall effect of light~\cite{hall1,hall2}. This idea has also been used to amplify small transverse deflections and frequency changes of optical beams~\citep{def1,def2,def3}. Because of the importance of weak measurement in applications, there has been much research work on this issue~\cite{w1,w2,w3,w4,w5,w6,w7,w8}.

In the seminal paper~\cite{aav}, Aharonov, Albert, and Vaidman (AAV) introduced the weak value
\begin{equation}
A_w=\frac{\bra{\psi_f} A \ket{\psi_i}}{\langle \psi_f | \psi_i \rangle}
\end{equation}
of an observable $A$, where $\ket{\psi_i}$ and $\ket{\psi_f}$ are the preselected and postselected states (PPS), respectively. For the interaction Hamiltonian $H=-g\delta(t-t_0)A \otimes q$, and the weak value $A_{w}=a+ib$ (where $a, b \in \mathcal{R}$), Jozsa~\cite{jozsa} has derived that the final mean pointer momentum and position satisfy
\begin{equation}\label{eq:jozsa}
\begin{split}
{\langle p \rangle}_f={\langle p \rangle}_i + ga-gb(m\frac{d}{dt}\mathrm{Var}_q), \\
{\langle q \rangle}_f={\langle q \rangle}_i-2gb(\mathrm{Var}_q),
\end{split}
\end{equation}
where ${\langle q \rangle}_i$ and ${\langle p \rangle}_i$ are the original mean pointer position and momentum, $\mathrm{Var}_q$ is the variance of position in the initial pointer state, and $m$ is the mass of the pointer. It seems that for $\langle \psi_f | \psi_i \rangle \to 0$, the shifts of the pointer's position and momentum might be arbitrarily large. In fact arbitrarily large shifts cannot be obtained, as there is one approximation step in the derivation \cite{jozsa}:
\begin{equation}\begin{split}
\ket{\Phi'} &=\bra{\psi_f}e^{igAq}\ket{\psi_i}\ket{\Phi} \\
             &\approx \langle \psi_f | \psi_i \rangle (I+igA_wq) \ket{\Phi},
\end{split}\end{equation}
where $\ket{\Phi}$ and $\ket{\Phi'}$ are the original and final states of the measuring device. As pointed out in~\cite{prd}, the validity of the approximation in the above equation requires that
\begin{equation}
|g^n q^n\langle \psi_f | A^n | \psi_i \rangle| \ll |\langle \psi_f | \psi_i\rangle|,
\end{equation}
for all $n \geq 2$. When $\langle \psi_f | \psi_i \rangle \to 0$, the above requirement is not satisfied, so the measurement results given by Eq. ($\ref{eq:jozsa}$) are not valid. Thus, we need some more precise descriptions of weak measurements. There has been some research on weak measurements beyond AAV's formalism~\cite{with1,with2}. By retaining the second order terms of $g$, the formalism of weak measurements is generalized to the general preselected and postselected states in~\cite{with2}. In that reference, it was shown that the weak measurement outcomes cannot increase arbitrarily by decreasing the overlap between the PPS.

In this paper, we calculate the shifts of the pointer position and momentum without any approximation. The interaction between the system and measuring device is not assumed to be very weak. So, the expressions we obtained are not only valid for the weak interaction cases, but also legitimate for the strong interaction cases. We also give the maximum shifts of the measuring device for the two-dimensional systems by choosing appropriate PPS. Using weak measurement, the improvements of the signal-to-noise ratio (SNR) and the enhancements of the measure sensitivity (MS) are discussed in the paper.
When we consider the probabilities reduced by postselection, we find the SNR and MS cannot be effectively improved in weak measurement. We also show that the SNR and MS can be effectively improved without considering the probabilities.

This paper is organized as follows. In Sec. II we will give the general precise expressions of the pointer shifts. The maximum pointer shifts with a two-dimensional system and the explicit results for the Stern-Gerlach experiment are derived in Sec. III.  Section IV is devoted to studying the improvement of the SNR brought by a weak measurement. The discussion of the enhancements of the MS is given in Sec. V. A short conclusion is presented in Sec. VI.

\section{Precise expressions of the measurement results}
In this section, we derive the precise expressions of the measurement results with preselection and postselection, without any restriction on the strength of the interaction between the system and the measuring device. The preselected and postselected states of the quantum system are denoted by $\ket{\psi_i}$ and $|\psi_f\rangle$ respectively, and the initial state of the measuring device is denoted by $\ket{\Phi}$. The interaction Hamiltonian between the system and the device is described as
\begin{equation}\label{x1}
H=-g\delta(t-t_0)A\otimes D,
\end{equation}
where $g$ is the coupling strength with $g > 0$, $A$ is an observable of the quantum system, and $D$ is a variable of the measuring device. Throughout this article, for the purpose of convenience, by $A \to A/\parallel A \parallel$, and $g \to g\parallel A \parallel$, we redefine $A$ and $g$ such that $A$ becomes a dimensionless operator with a unit norm. It is assumed that the complete orthonormal eigenstates of the observable $A$ are $\{\ket{a_m}\}$, the corresponding eigenvalues are $\{a_m\}$. So, the preselected state $\ket{\psi_i}$ can be written as
\begin{equation}
\ket{\psi_i}=\sum_m \alpha_m \ket{a_m}.
\end{equation}
And the overall state of the system and measuring device is
\begin{equation}\label{eq:c1}
\ket{\Psi}=\sum_m \alpha_m \ket{a_m}\ket{\Phi}.
\end{equation}
After the interaction described by Eq. (\ref{x1}) the overall state evolves into
\begin{equation}\label{eq:c2}
\ket{\Psi'}=e^{-i\int H \mathrm{d} t}\ket{\Psi}=\sum_m \alpha_m e^{iga_m D}\ket{a_m}\ket{\Phi},
\end{equation}
with $\hbar=1$ throughout this paper. The postselected state can also be written as $\ket{\psi_f}=\sum_m \beta_m \ket{a_m}$. After postselection, the final device state is
\begin{equation}\label{eq:c3}
\ket{\Phi'}=\langle \psi_f |\Psi'\rangle=\sum_m \alpha_m \beta^{*}_m e^{iga_m D}\ket{\Phi},
\end{equation}
which is not a normalized state. The probability of obtaining the final state $\ket{\Phi'}$ is
\begin{equation}\label{eq:c4}
P=\langle \Phi'| \Phi'\rangle = \sum_{mn}\alpha_m \beta^{*}_m \alpha^{*}_n \beta_n \bra{\Phi} e^{ig(a_m-a_n)D}\ket{\Phi}.
\end{equation}
The expectation value of a pointer observable $M$ is given by
\begin{equation}\label{eq:c5}\begin{split}
\langle M' \rangle &=\frac{\langle \Phi'| M |\Phi' \rangle}{\langle \Phi'| \Phi'\rangle }\\
&=\frac {\sum_{mn}\gamma_m \gamma^{*}_n \bra{\Phi} e^{-iga_nD} M e^{iga_mD}\ket{\Phi}}{\sum_{mn} \gamma_m \gamma^{*}_n \bra{\Phi} e^{ig(a_m-a_n)D}\ket{\Phi}},
\end{split}\end{equation}
where $\gamma_m= \alpha_m \beta^{*}_m$. Thus, we have given the general and precise expression of the device's readings.

Without loss of generality, the initial state of the pointer device is assumed to be a Gaussian wave function centered on $q=0$ and $p=0$,
\begin{equation}\begin{split}
&\Phi(q)=\frac{1}{ (2 \pi\Delta^{2})^{\frac{1}{4}}}\exp({-\frac{q^2}{4\Delta^2}}),\\
&\Phi(p)=\frac{(2\Delta^{2} )^{\frac{1}{4}}}{\pi^{\frac{1}{4}}}\exp({-\Delta^2 p^2}),
\end{split}\end{equation}
where $q$ and $p$ are the position and momentum variables, and the standard deviations of them are $\Delta q= \Delta$ and $\Delta p= \frac{1}{2\Delta}$. In this paper, we consider $D=q$, and the interaction $H=-g\delta(t-t_0)A\otimes q$. From Eq. (\ref{eq:c3}), the final state of the pointer device after postselection is
\begin{equation}\label{eq:c6}
\ket{\Phi'}={1}/{ (2 \pi\Delta^{2})^{{1}/{4}}}\sum_m \gamma_m e^{iga_mq}e^{{-{q^2}/{4\Delta^2}}},
\end{equation}
and the probability of obtaining the final state $\ket{\Phi'}$ is
\begin{equation}\label{eq:c7}
P=\langle \Phi'| \Phi'\rangle = \sum_{mn}\gamma_m \gamma^{*}_n e^{-{\Delta^2 g^2 (a_m-a_n)^2}/2}.
\end{equation}
From Eq. ($\ref{eq:c5}$), we can directly calculate the expectation values of the pointer momentum and position
\begin{equation}\label{eq:c8} \begin{split}
\langle p' \rangle = \frac{\frac{g}{2} \sum_{mn}\gamma_m \gamma^{*}_n (a_m+a_n)e^{-{\Delta^2 g^2 (a_m-a_n)^2}/2}}{\sum_{mn}\gamma_m \gamma^{*}_n e^{-{\Delta^2 g^2 (a_m-a_n)^2}/2}},
\end{split}\end{equation}
and
\begin{equation}\label{eq:c9}\begin{split}
\langle q' \rangle = \frac{i g \Delta^2 \sum_{mn}\gamma_m \gamma^{*}_n (a_m-a_n)e^{-{\Delta^2 g^2 (a_m-a_n)^2}/2}}{\sum_{mn}\gamma_m \gamma^{*}_n e^{-{\Delta^2 g^2 (a_m-a_n)^2}/2}}.
\end{split}\end{equation}
As the expectation values of the pointer momentum and position in the initial state are both zero, the values given in Eqs. ($\ref{eq:c8}$) and ($\ref{eq:c9}$) are also the average shifts of the pointer momentum and position, which are denoted by $\delta p'$ and $\delta q'$, respectively.

In the rest of this section, we discuss two extreme cases in terms of the interaction strength: the weak interaction case and the strong interaction case.

First, we discuss the weak interaction case (i.e., the coupling strength is weak $g \ll 1/{2\Delta} =\Delta p$). We have $e^{-{\Delta^2 g^2 (a_m-a_n)^2}/2} \approx 1$, and
\begin{equation}\label{eq:c10}\begin{split}
\delta p' &\approx \frac{\frac{g}{2} \sum_{mn}\gamma_m \gamma^{*}_n (a_m+a_n)}{\sum_{mn}\gamma_m \gamma^{*}_n} \\
                  &= \frac{g}{2}(\frac{\bra{\psi_f}A \ket{\psi_i}}{\langle \psi_f | \psi_i \rangle}+\frac{\bra{\psi_i} A \ket{\psi_f}}{\langle \psi_i | \psi_f \rangle})= g \mathrm{Re}(A_w),
\end{split}
\end{equation}
and
\begin{equation}\label{eq:c11}\begin{split}
\delta q' &\approx \frac{ig {\Delta}^2 \sum_{mn}\gamma_m \gamma^{*}_n (a_m-a_n)}{\sum_{mn}\gamma_m \gamma^{*}_n} \\
                  &= ig {\Delta}^2 (\frac{\bra{\psi_f} A \ket{\psi_i}}{\langle \psi_f | \psi_i \rangle}-\frac{\bra{\psi_i} A \ket{\psi_f}}{\langle \psi_i | \psi_f \rangle})= -2g (\Delta q)^2 \mathrm{Im}(A_w).
\end{split}
\end{equation}
If the real part of the weak value $A_w$ is big, the shift $\delta p'$ can be much larger than the shift of the pointer without postselection. This amplification effect is implemented in experiments~\cite{exp1,exp2}. The results in Eqs. (\ref{eq:c10}) and (\ref{eq:c11}) are the same as those given in Eq. (\ref{eq:jozsa}).
However, it should be pointed out that the approximation step may not be legitimate when $\langle \psi_f | \psi_i \rangle \to 0$. Therefore, in a general discussion without restriction on the overlap between preselection and postselection states,
we use Eqs. ($\ref{eq:c8}$) and ($\ref{eq:c9}$) instead.

Next, we discuss the strong interaction case (i.e., $g \gg \Delta p$). We have $e^{-{\Delta^2 g^2 (a_m-a_n)^2}/2} \approx 0$ for $a_m \neq a_n$, and then $\delta q' \approx 0$. If the eigenvalues of $A$ are nondegenerate, the shift of the pointer momentum can be approximately expressed as
\begin{equation}\label{eq:c12}\begin{split}
\delta p' & \approx \frac{g\sum_{m}a_m |\gamma_m|^2}{\sum_{m} |\gamma_m|^2}= \frac{g \mathrm{Tr}(A \proj{\psi_f} \rho_i) }{\mathrm{Tr}(\proj{\psi_f}\rho_i)},
\end{split}\end{equation}
where $\rho_i=\sum_m |{\langle a_m | \psi_i\rangle}|^2 \proj{a_m}$ is the state of the system after projection onto the eigenstates $\{\ket{a_m}\}$. When the eigenvalues are degenerate, let the eigenvalue $a_s$ be $d_s$-fold degenerate, we can re-label the eigenstates of $A$ as $\{\ket{a_s^{k_s}}\}$, where $k_s =1,2,. . ., d_s$. Then the PPS can be rewritten as $\ket{\psi_i}=\sum_{s,k_s} \alpha_{s}^{k_s}\ket{a_s^{k_s}}$ and $\ket{\psi_f}=\sum_{s,k_s} \beta_{s}^{k_s}\ket{a_s^{k_s}}$. From Eq. (\ref{eq:c8}) and $e^{-{\Delta^2 g^2 (a_m-a_n)^2}/2} \approx 0$ for $a_{m} \neq a_n$, we get
\begin{equation}\label{e1}
\delta p' \approx \frac{g\sum_s|\sum_{k_s=1}^{d_s} \gamma_{s}^{k_s}|^2a_{s}}{\sum_s|\sum_{k_s=1}^{d_s} \gamma_{s}^{k_s}|^2},
\end{equation}
where $\gamma_{s}^{k_s}= \alpha_{s}^{k_s}\beta_{s}^{k_s*}$. The shifts given by Eqs. (\ref{eq:c12}) and (\ref{e1}) are the results for a strong measurement, and they can be regarded as the ideal quantum measurement results which are independent of the device's states. So, we have got the ideal measurement results by increasing the strength of the interaction between the quantum system and measuring device. From the Eq. (\ref{eq:c12}) and Eq. (\ref{e1}), we can get that
\begin{equation}\label{eq:c13}
a_{\min}g \leq \delta p' \leq a_{\max}g,
\end{equation}
where $a_{\min}$ and $a_{\max}$ are the minimum and maximum eigenvalues of the observable $A$, respectively. The measurement results given in Eq. (\ref{eq:c13}) mean that the amplification effect disappears when the interaction strength is strong.

\section{Maximum pointer shifts}
In this section, we derive the maximum shifts of the pointer position and momentum for a measurement with preselection and postselection on a qubit system. The interaction between the system and the measuring device is described by the Hamiltonian $H=-g\delta(t-t_0)A \otimes q$, which is not assumed to be weak unless specified. As the system to be measured is two-dimensional, so the orthonormal eigenstates of the observable $A$ can be denoted as $\{\ket{0},\ket{1}\}$, and the corresponding eigenvalues are denoted as $\{a_1,a_2\}$, respectively. Without loss of generality, we assume $a_1 \geq a_2$. Then the PPS $\ket{\psi_i}$ and $\ket{\psi_f}$ can be written as
\begin{equation}\label{eq:c14}\begin{split}
\ket{\psi_i}=\cos \frac{\theta_1}{2}\ket{0}+\sin \frac{\theta_1}{2} e^{i \phi_1}\ket{1},\\
\ket{\psi_f}=\cos \frac{\theta_2}{2}\ket{0}+\sin \frac{\theta_2}{2} e^{i \phi_2}\ket{1},
\end{split}\end{equation}
where $\theta_1, \theta_2 \in [0,\pi)$ and $\phi_1, \phi_2 \in [0,2\pi)$. From Eqs. ($\ref{eq:c8}$) and ($\ref{eq:c9}$), we get
\begin{equation}\label{eq:c15}\begin{split}
\delta p' = \frac {a_1+a_2}{2}g + \frac{g(a_1-a_2)(\cos{\theta_1}+\cos{\theta_2})}{2N(\theta_1,\theta_2, \phi_1, \phi_2)},
\end{split}\end{equation}
and
\begin{equation}\label{eq:c16}
\delta q' = \frac{g\Delta^2(a_1-a_2)\sin{\theta_1}\sin{\theta_2}\sin{(\phi_1-\phi_2)}e^{-g^2\Delta^2(a_1-a_2)^2/2}}{N(\theta_1,\theta_2, \phi_1, \phi_2)},
\end{equation}
where $N(\theta_1,\theta_2, \phi_1, \phi_2)=1+\cos{\theta_1}\cos{\theta_2}+\sin{\theta_1}\sin{\theta_2}\cos{(\phi_1-\phi_2)}e^{-g^2\Delta^2(a_1-a_2)^2/2}$. When the two states $\ket{\psi_i}$ and $ \ket{\psi_f}$ are orthogonal to each other, we obtain that
\begin{equation}\label{eq:c17}
\delta p' =\frac{a_1+a_2}{2}g, \quad \delta q' =0,
\end{equation}
while AAV's formalism cannot give any measurement result for two orthogonal PPS.

Now, for a given pointer state and coupling strength $g$, we search for the extreme values of $\delta p'$ and $\delta q' $ over all the PPS. Let $\phi_0=\phi_1-\phi_2$, from $\frac{\partial \delta p' }{\partial \phi_0}=0$, we get $\sin \phi_0=0$, and
\begin{equation}\label{eq:c18}
\delta p' =\frac {a_1+a_2}{2}g + \frac{g(a_1-a_2)t}{1+t^2\pm (1-t^2)e^{-g^2\Delta^2(a_1-a_2)^2/2}},
\end{equation}
where $t=\cos{\frac{(\theta_1+\theta_2)}{2}}/\cos{\frac{(\theta_1-\theta_2)}{2}}$. By $\frac{\partial \delta p'}{\partial t}=0$, we get the extreme values of the shift of the pointer momentum
\begin{equation}\label{eq:c19}
\delta p'= \frac{a_1+a_2}{2}g\pm \frac{a_1-a_2}{2\sqrt{1-e^{-g^2 \Delta^2 (a_1-a_2)^2}}}g.
\end{equation}
As $a_1 \geq a_2$, it is obtained that
\begin{equation}\label{eq:c20}\begin{split}
\delta p'_{\min}= \frac{a_1+a_2}{2}g - \frac{a_1-a_2}{2\sqrt{1-e^{-g^2 \Delta^2 (a_1-a_2)^2}}}g,\\
\delta p'_{\max}= \frac{a_1+a_2}{2}g + \frac{a_1-a_2}{2\sqrt{1-e^{-g^2 \Delta^2 (a_1-a_2)^2}}}g.
\end{split}\end{equation}
Let $M=1+e^{-g^2 \Delta^2 (a_1-a_2)^2/2}$ and $W=1-e^{-g^2 \Delta^2 (a_1-a_2)^2/2}$. When $\phi_1-\phi_2=\pm \pi $ and $t=-\sqrt{W/M}$, $\delta p'$ reaches its minimum value $\delta p'_{\min}$; when $\phi_1-\phi_2=\pm \pi$ and $t=\sqrt{W/V}$, $\delta p'$ reaches its maximum value $\delta p'_{\max}$, where $t=\cos{\frac{(\theta_1+\theta_2)}{2}}/\cos{\frac{(\theta_1-\theta_2)}{2}}$.
For the weak interaction case $g \ll \frac{1}{2\Delta} =\Delta p$, we have $\pm\sqrt{W/M} \to 0$, the extreme shift is achieved when $\theta_1+\theta_2 \approx \pi$.  It is obvious that $\phi_1-\phi_2=\pm \pi $ and $\theta_1+\theta_2 \approx \pi$ imply $\langle \psi_f | \psi_i \rangle \to 0$. So for the weak interaction cases, the shift of the pointer reaches its extreme value when the PPS are approximatively orthogonal.

Furthermore, for the weak interaction cases with $g \ll \Delta p$, we get $\delta p'_{\min} \approx \frac{a_1+a_2}{2}g-\frac{1}{2\Delta}\approx -\Delta p$ and $\delta p'_{\max} \approx \Delta p$ which only depend on the original state of the measuring device. For the strong interaction cases with $g \gg \Delta p$, we get $\delta p'_{\min} \approx a_2g$, and $\delta p'_{\max} \approx a_1g$, which depend on the eigenvalues of $A$ and the coupling strength $g$.

By a similar derivation, we obtain that
\begin{equation}\label{eq:c21}\begin{split}
\delta q'_{\min}= - \frac{g\Delta^2 (a_1-a_2)e^{-\Delta^2g^2(a_1-a_2)^2/2}}{\sqrt{1-e^{-g^2 \Delta^2 (a_1-a_2)^2}}},\\
\delta q'_{\max}= \frac{g\Delta^2 (a_1-a_2)e^{-\Delta^2g^2(a_1-a_2)^2/2}}{\sqrt{1-e^{-g^2 \Delta^2 (a_1-a_2)^2}}}.
\end{split}\end{equation}
The minimum shift $\delta q'_{\min}$ of the pointer position is achieved when $\theta_1+\theta_2=\pi$ and $\phi_1-\phi_2=\pi+\arccos e^{-g^2 \Delta^2 (a_1-a_2)^2/2}$;
while the maximum shift $\delta q'_{\max}$ is achieved when $\theta_1+\theta_2=\pi$ and $\phi_1-\phi_2=\pi-\arccos e^{-g^2 \Delta^2 (a_1-a_2)^2/2}$. For $g \ll \Delta p$, $\pi\pm \arccos e^{-g^2 \Delta^2 (a_1-a_2)^2/2} \to \pi$, so the extreme shift of the pointer position is achieved only when $\langle \psi_f | \psi_i \rangle \to 0$.  For the weak interaction cases $g \ll \Delta p$, the minimum and maximum shifts of the pointer position are given by $\delta q'_{\min} \approx -\Delta = -\Delta q$, and $\delta q' _{\max} \approx \Delta = \Delta q$; while for the strong interaction cases $g \gg \Delta p$, we get $\delta q'_{\min} \approx \delta q'_{\max} \approx 0$.

Here, we have obtained the extreme values of the pointer's shifts for all PPS. When $g \ll \Delta p$, we find that the maximum shifts are independent of the coupling strength $g$, and by choosing the appropriate PPS, the maximum values of pointer shifts $\delta p'=\Delta p$ and $\delta q'=\Delta q$ can be obtained. So for the weak interaction case, due to the postselection, the signal is amplified by a factor of $\frac{1}{2|g|\Delta}$ which could be very large. When $g \gg \Delta p$, it is obtained that $a_2g \leq \delta p' \leq a_1 g$ and $\delta q' =0$, and these results mean that there is no amplification effect at all under the strong interaction circumstance.

Now we consider the Stern-Gerlach experiment which is widely discussed in weak measurements~\cite{aav,with2,Lorenzo}. In this experiment, spin-${1}/{2}$ particles with the spin pointed in the direction $\bm{n} (\theta,\phi)$ travel through an inhomogeneous (in the $z$ direction) magnetic field, and then are postselected in the $x$ direction. So, we have $\ket{\psi_i}=\cos{\frac{\theta}{2}\ket{0}+\sin{\frac{\theta}{2}}}e^{i\phi}\ket{1}$ and $\ket{\psi_f}=\frac{1}{\sqrt{2}}(\ket{0}+\ket{1})$. The initial pointer state is $\ket{\Phi}=\Delta^{1/2}(2 \pi )^{-1/4}e^{-z^2/4}f(x,y)$, where $f(x,y)$ is a normalized wave function. The interaction Hamiltonian can be written as $H=-g\delta(t-t_0)\sigma_z \otimes z$, where $g=\mu\frac{\partial B_z}{\partial z}$ depends on the magnetic field. After postselection, the final pointer state is
\begin{equation}\label{eq:b00}
\ket{\Phi'}=(\frac{\Delta^2}{8 \pi })^{1/4}(\cos{\frac{\theta}{2}e^{igz}+\sin{\frac{\theta}{2}}}e^{-i\phi}e^{-igz})e^{-z^2/4}f(x,y).
\end{equation}
Substituting $\ket{\Phi'}$ into Eq. ($\ref{eq:c7}$), and from Eqs. ($\ref{eq:c8}$) and ($\ref{eq:c9}$), we get that
\begin{equation}\label{eq:c22}\begin{split}
\delta p'_z = \frac{g\cos{\theta}}{1+\sin{\theta}\cos{\phi}e^{-2\Delta^2 g^2}},\\
\delta z' = \frac{2g\Delta^2 \sin{\theta}\sin{\phi}e^{-2\Delta^2g^2}}{1+\sin{\theta}\cos{\phi}e^{-2\Delta^2 g^2}},\\
P = \frac{1+\sin{\theta}\cos{\phi}e^{-2\Delta^2 g^2}}{2}.
\end{split}\end{equation}

\begin{figure}[t]
\centering \includegraphics[scale=0.25]{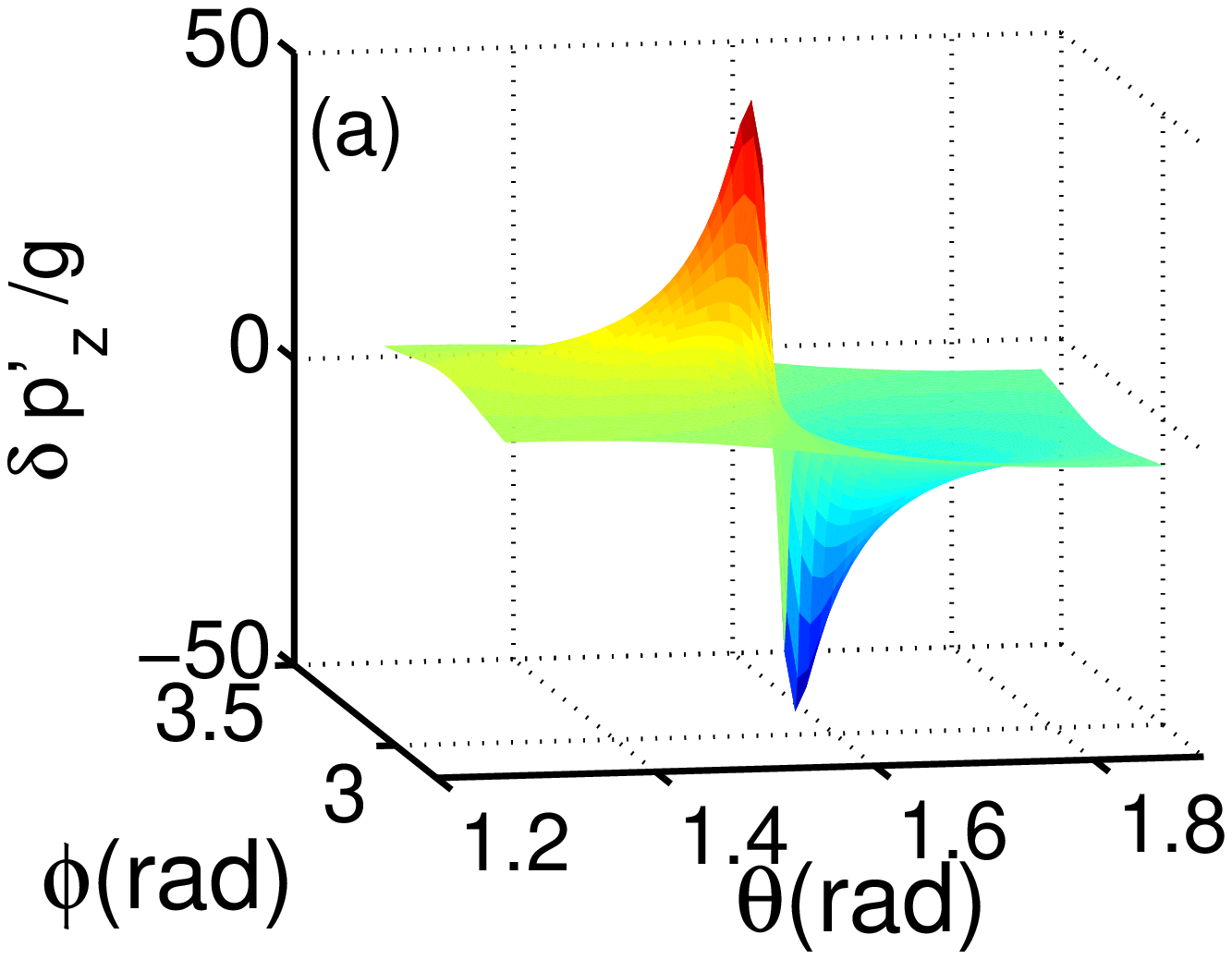} \includegraphics[scale=0.25]{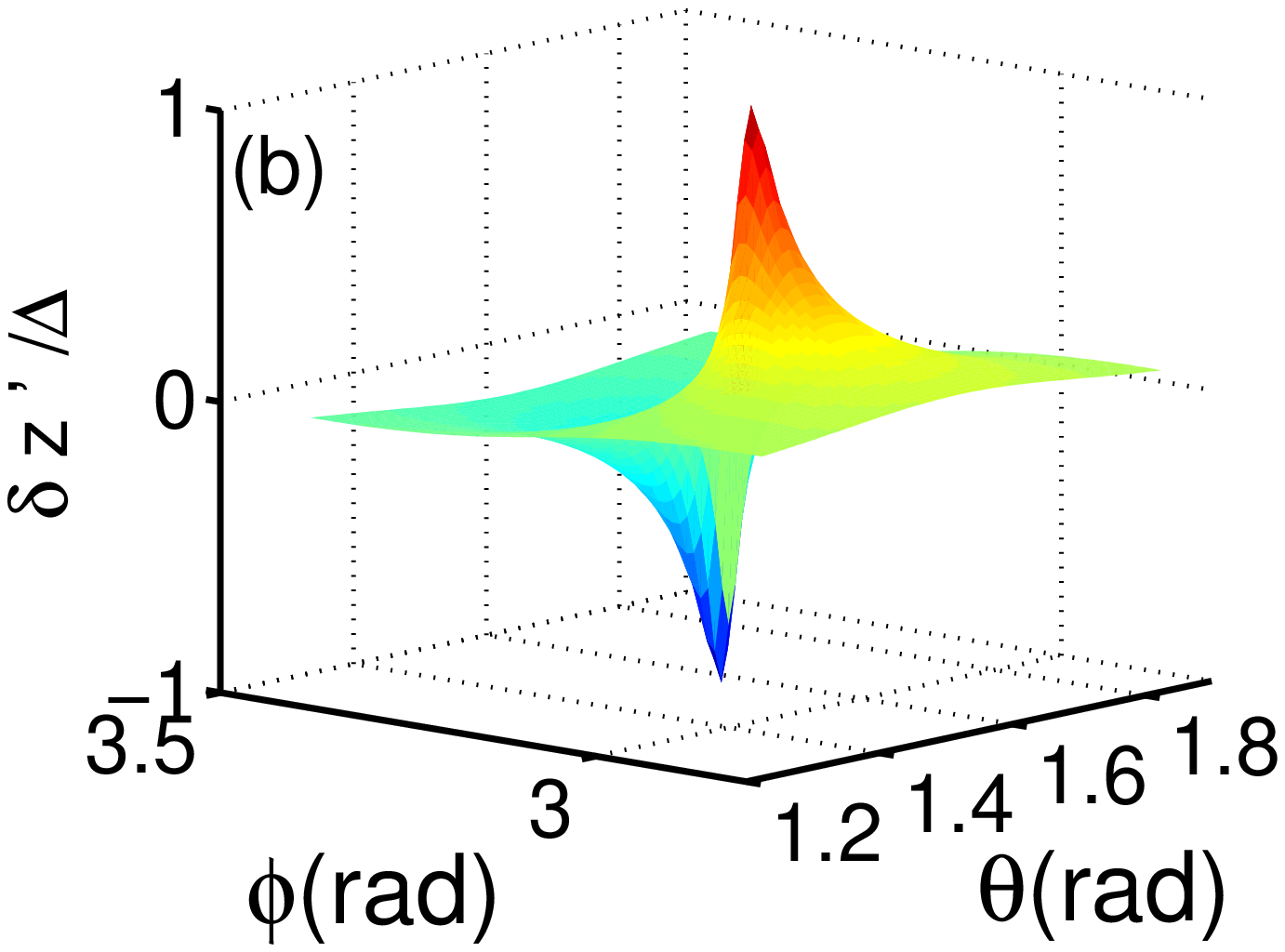} \caption{(Color online)The shifts of the pointer momentum and
position with $g=0.02\Delta p_z$. The extremum values are obtained near $\theta\approx \pi/2$ and $\phi\approx \pi$.}
\label{fig:1}
\end{figure}

In Fig. 1, the shifts are depicted for $g=0.02\Delta p_z$. It can be seen that, the maximal pointer shifts are $\delta p'_{z,\max} \approx 50g=1/{2\Delta}=\Delta p_z$ and $\delta z'_{\max}\approx \Delta = \Delta z $ which are the extreme values obtained in Eqs. ($\ref{eq:c20}$) and ($\ref{eq:c21}$). The maximum amplification factor obtained here is $50$, which can not be increased anymore by decreasing the overlap between the PPS (it can never reach $100$ as in~\cite{aav}).

From $\frac{\partial \delta p'_z}{\partial \theta}=0$ and $\frac{\partial \delta p'_z}{\partial \phi}=0$, we get the maximal shift of the pointer momentum
\begin{equation}\label{eq:2301}\begin{split}
\delta p'_{z,\max}=\frac{g}{\sqrt{1-e^{-4\Delta^2g^2}}}.
\end{split}\end{equation}
The maximum is achieved when $\theta=\theta_{\mathrm{opt}}=\arcsin e^{-2\Delta^2g^2}$ and $\phi=\pi$. For $g \ll \Delta p_z$, $\theta_{\mathrm{opt}}=\arcsin e^{-2\Delta^2g^2} \approx \pi/2$, we have $\langle \psi_f|\psi_i\rangle \to 0$. It means the optimum PPS to achieve the maximum shift of the pointer momentum are approximatively orthogonal. And the probability of obtaining this maximum value is given by $P_{p_z,\max}=\frac{1-e^{-4\Delta^2g^2}}{2}$.

From $\frac{\partial \delta z'}{\partial \theta}=0$ and $\frac{\partial \delta z'}{\partial \phi}=0$, we get the maximal shift of the pointer position
\begin{equation}\label{eq:2302}
\delta z'_{\max}=\frac{2g\Delta^2e^{-2\Delta^2g^2}}{\sqrt{1-e^{-4\Delta^2g^2}}},
\end{equation}
which can be achieved when $\theta=\pi/2$ and $\phi= \pi-\arccos e^{-2\Delta^2g^2}$. For $g \ll \Delta p_z$, $\phi= \pi-\arccos e^{-2\Delta^2g^2} \approx \pi$, and we have $\langle \psi_f|\psi_i\rangle \to 0$. The probability of obtaining the maximum shift is $P_{z,\max}=\frac{1-e^{-4\Delta^2g^2}}{2}$. As $P_{p_z,\max}=P_{z,\max}$, we relabel the probability $P_{\max}=P_{p_z,\max}=P_{z,\max}=\frac{1-e^{-4\Delta^2g^2}}{2}$.

In Fig. 2, the maximal pointer shifts and the probability are pictured with different coupling strength $g$. It can be seen from Figs. 2(a) and 2(b) that when $g \ll \Delta p_z$, the maximum shifts $\delta p'_z $ and $\delta z'$ are given by $\Delta p_z$ and $\Delta z$ respectively. However, the probability of obtaining the maximal shifts is very small, $P_{\max} \approx 2\Delta^2g^2$. When $g \gg \Delta p_z$, we have $\delta p'_{z,\max}=g$ and $\delta z'_{\max}=0$, which is expected from a strong measurement.

\begin{figure}[t]
\centering \includegraphics[scale=0.25]{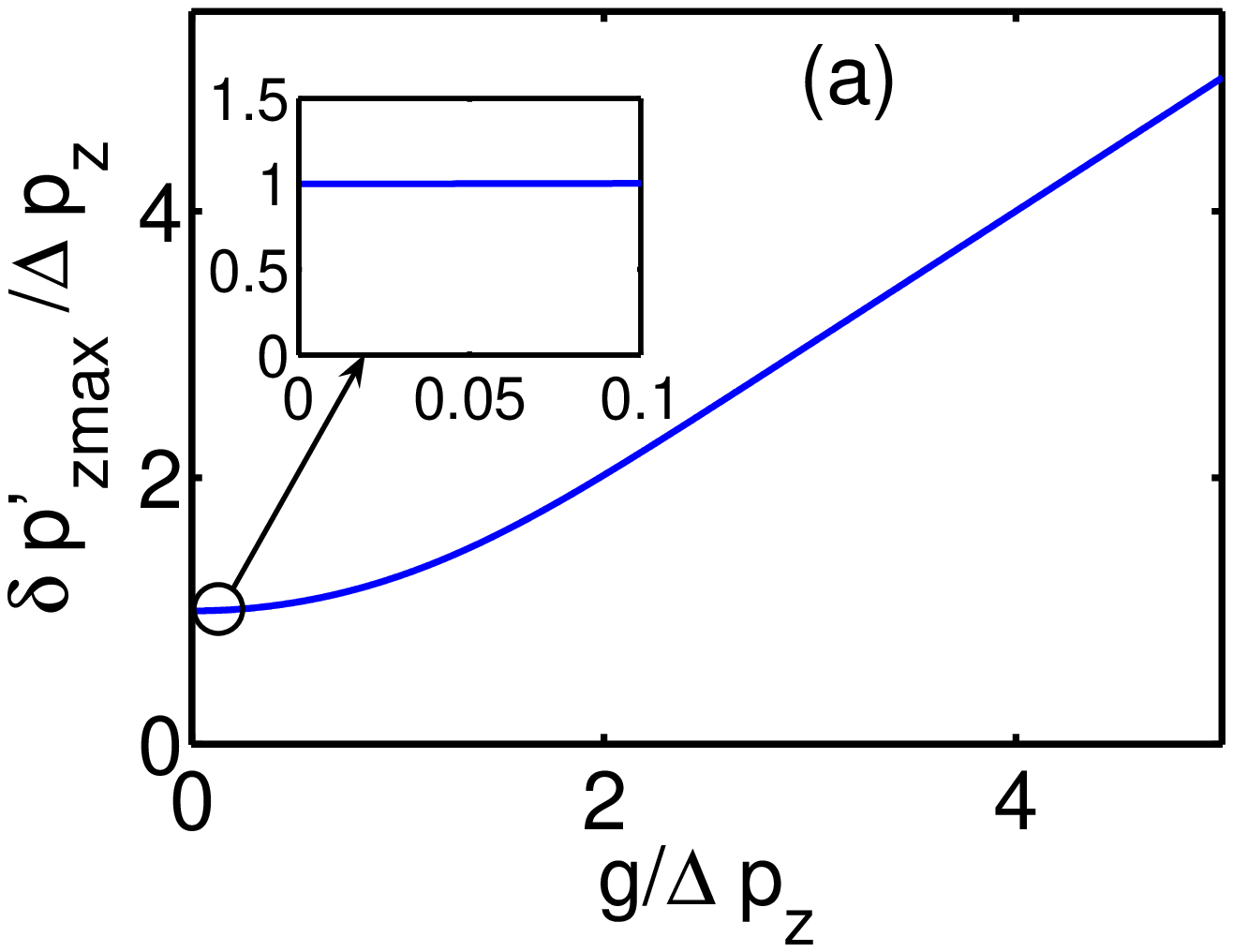} \includegraphics[scale=0.25]{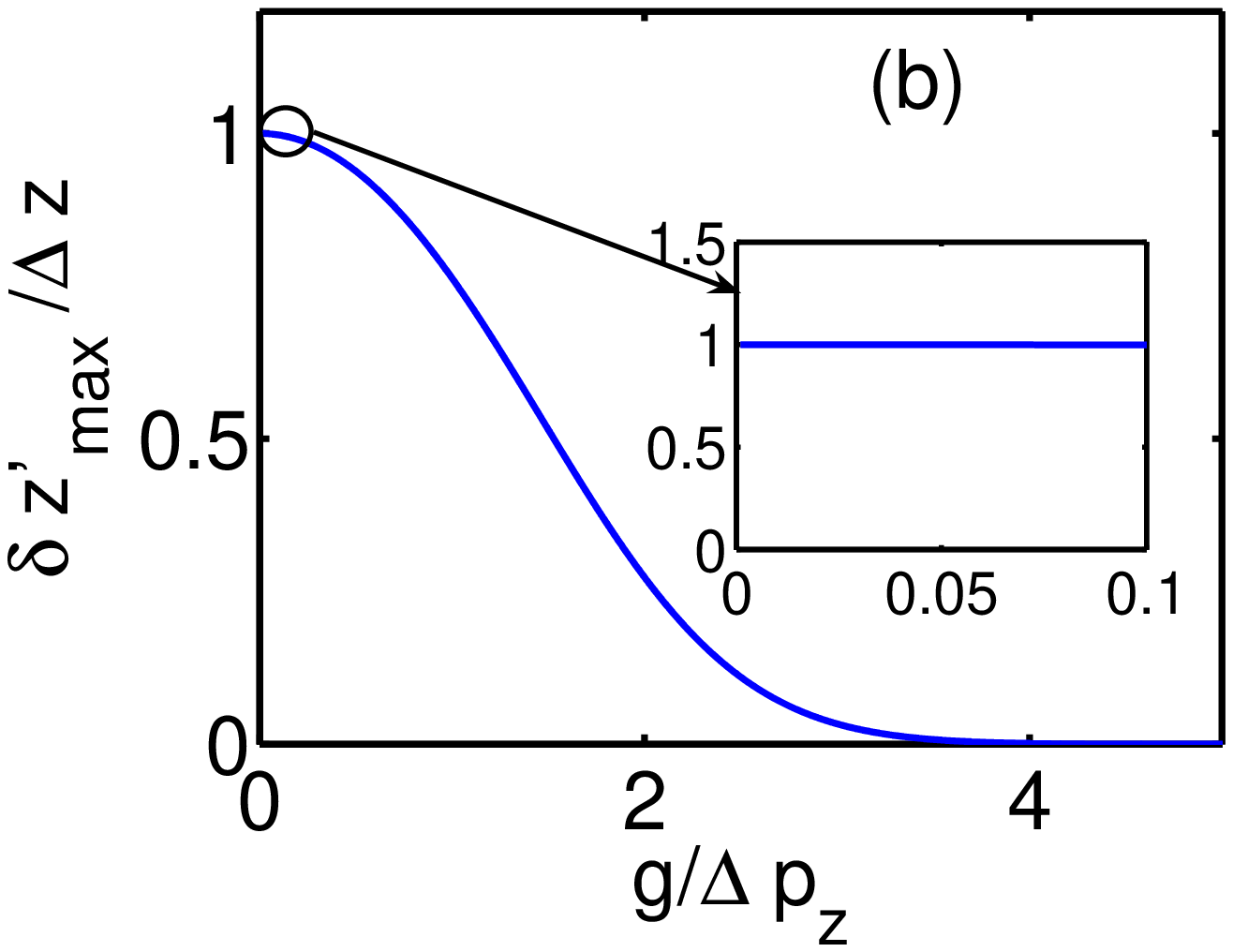} \includegraphics[scale=0.25]{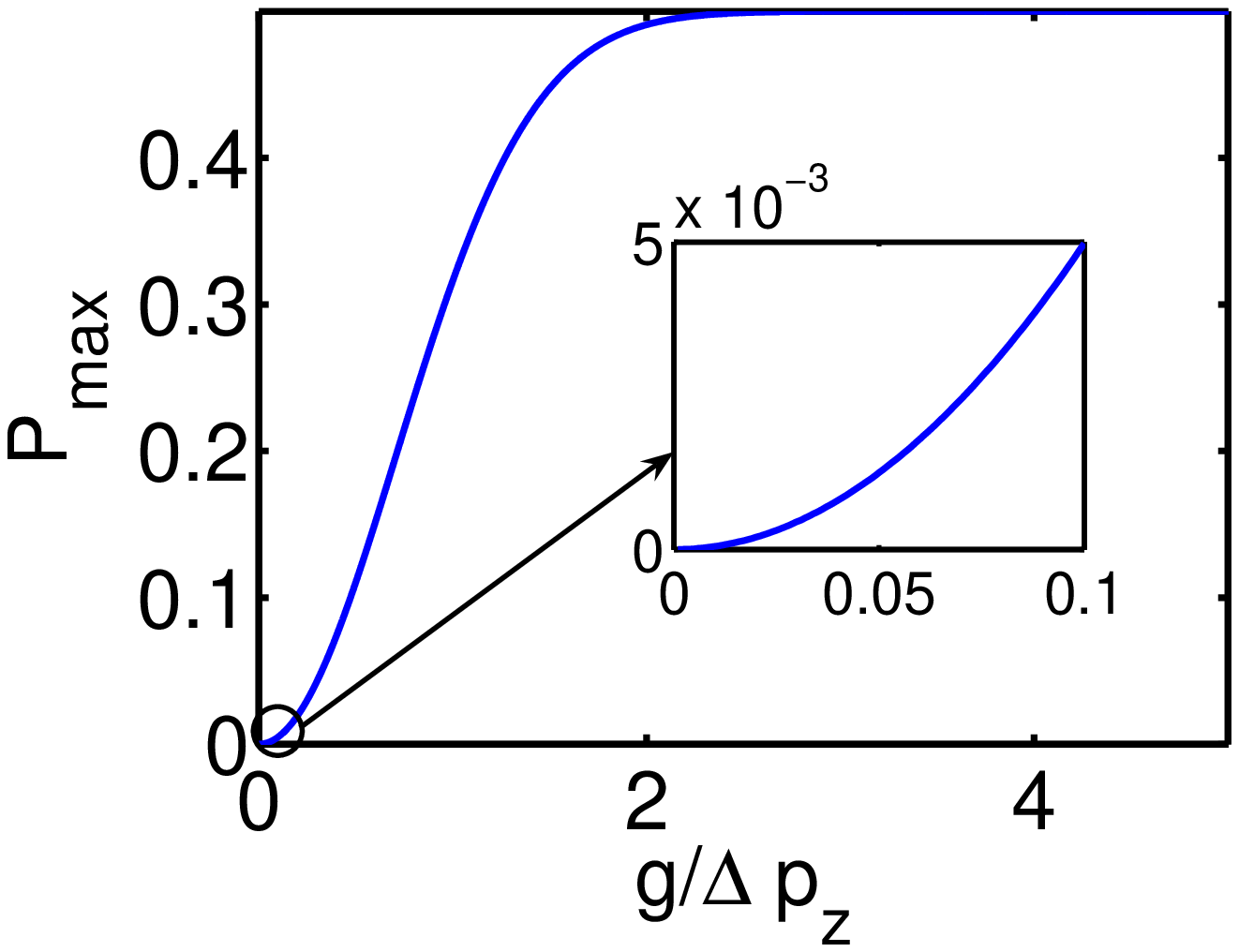} \caption{(Color online) The maximal shifts of the pointer momentum and position, and the probabilities of obtaining the maximum. When $g \ll \Delta p_z$, we have $\delta p'_{z,\max} \approx \Delta p_z$ and $\delta z'_{\max} \approx \Delta z$; when $g \gg \Delta p_z$, we have $\delta p'_{z,\max} \approx g$ and $\delta z'_{\max} \approx 0$, which are the strong measurement results. When $g \ll \Delta p_z$, the probability to obtain the maximum shifts $P_{\max} \approx 2\Delta^2g^2$ is very small.}
\label{fig:2}
\end{figure}

\section{Improvement of the signal-to-noise ratio via weak measurements}
In this section, we study the improvements of the signal-to-noise ratio (SNR) by weak measurements. SNR has been studied in many reference~\cite{def2,snr1,FXS2011,WZ11}; it was claimed that the SNR can be largely improved by the weak measurement processes. Here, we show that the SNR can be improved for some practical cases when the probability decrease due to postselection is not considered. However, we also show that the SNR can not be effectively improved when the probability decrease due to postselection is considered.

\subsection{No effective improvement of the SNR when the probabilities reduced by postselection are considered}
The signal considered here is the average shift of the pointer variable $X$, and the noise is represented by the standard deviation $\sigma_{X}$. If the number of the measurement repetition is $N$, as a result of the central limit theorem, the mean standard deviation is $\overline{\sigma_{X}}=\frac{\sigma_{X}}{\sqrt{N}}$. Then we get the SNR
\begin{equation}\label{eq:d03}
R=\frac{|X|}{\overline{\sigma_{X}}}=\frac{\sqrt{N}|X|}{\sigma_{X}} .
\end{equation}
This definition is similar to the one given in~\cite{def2,snr1}.

Without postselection, the state of the measuring device after the interaction $H=-g\delta(t-t_0)A\otimes q$ is
\begin{equation}\label{eq:d01}\begin{split}
\rho_{D} &=\mathrm{Tr}_{S}(\ket{\Psi'}\bra{\Psi'})\\
          &=\sum_{m}|\alpha_m|^2 \ket{\Phi(p-a_mg)} \bra{\Phi(p-a_mg)},
\end{split}\end{equation}
where $\ket{\Psi'}$ is the composite state of the system and
measuring device given by Eq. (\ref{eq:c2}).  The shift of the pointer momentum without postselection is
\begin{equation}\label{eq:d02}
\delta p= \mathrm{Tr}(p\rho_{D})-p_0=\sum_m |\alpha_m|^2 a_mg,
\end{equation}
where $p_0$ is the original mean pointer momentum, and $\sum_m |\alpha_m|^2=1$. We get the SNR without postselection
\begin{equation}\label{eq:d04}
R_0=\frac{\sqrt{N}|\delta p|}{\Delta p}=\frac{\sqrt{N}|\sum_m |\alpha_m|^2 a_mg|}{\Delta p}.
\end{equation}
As $|\delta p |=|\sum_m |\alpha_m|^2 a_mg|\leq |a_{\max} g|$, where $a_{\max}$ is the eigenvalue of $A$ with the maximum absolute value, by choosing an appropriate input state we can get the optimal SNR without postselection is
\begin{equation}\label{eq:d05}
R_{0,\max}=\frac{\sqrt{N}|a_{\max} g|}{\Delta p}.
\end{equation}
The expectation values of $p$ and $q$ with postselection are given in Eqs. ($\ref{eq:c8}$) and ($\ref{eq:c9}$), respectively. The standard deviations of them are changed by postselection, and it can be obtained that
\begin{equation}\label{eq:d06}\begin{split}
\langle p'^2 \rangle & =\frac{\langle\Phi'| p^2 |\Phi'\rangle}{\langle \Phi'|\Phi'\rangle}\\
&=\frac{\frac{g^2}{4}\sum_{mn}\gamma_m \gamma^{*}_n (a_m+a_n)^2e^{-{\Delta^2 g^2 (a_m-a_n)^2}/2}}{\sum_{mn}\gamma_m \gamma^{*}_n e^{-{\Delta^2 g^2 (a_m-a_n)^2}/2}}+\frac{1}{4\Delta^2},
\end{split}\end{equation}
and
\begin{equation}\label{eq:d07}\begin{split}
\langle q'^2 \rangle &=\frac{\langle\Phi'| q^2| \Phi\rangle}{\langle \Phi'|\Phi'\rangle}\\
                  &= \frac{-g^2 \Delta^4 \sum_{mn}\gamma_m \gamma^{*}_n (a_m-a_n)^2e^{-{\Delta^2 g^2 (a_m-a_n)^2}/2}}{\sum_{mn}\gamma_m \gamma^{*}_n e^{-{\Delta^2 g^2 (a_m-a_n)^2}/2}}+\Delta^2,
\end{split}\end{equation}
where $|\Phi'\rangle$, the final device state after postselection,  is given by Eq. (\ref{eq:c6}).
So, we can get the standard deviations after postselection
\begin{equation}\label{eq:d08}
\Delta p'= \sqrt{\langle p'^2 \rangle -\langle p' \rangle ^2}, \Delta q'= \sqrt{\langle q'^2 \rangle -\langle q' \rangle ^2}.
\end{equation}
where $\langle p' \rangle$ and $\langle q' \rangle$ are given by Eqs. (\ref{eq:c8}) and (\ref{eq:c9}).
After postselection, the number of the measurement repetition is changed into $NP$, where $P$ is the probability of obtaining the postselected results. Then, we get the SNR with the postselection procedure
\begin{equation}\label{eq:d09}
R_p= \frac{\sqrt{N P}|\delta p'|}{\Delta p'}, R_q= \frac{\sqrt{NP}|\delta q'|}{\Delta q'}.
\end{equation}
Now we define the improvements of SNR by weak measurement as
\begin{equation}\label{eq:d10}\begin{split}
I_{p,I}=\frac{R_p}{R_{0,\max}}=\frac{\Delta p \sqrt{P}|\delta p'|}{\Delta p'|a_{\max}g|},\\
I_{q,I}=\frac{R_q}{R_{0,\max}}=\frac{\Delta p \sqrt{P} |\delta q'|}{\Delta q'|a_{\max}g|}.
\end{split}\end{equation}
They are referred to as type-I improvements of SNR, and in the definitions the probabilities reduced by postselection are taken into account. Next we will search the values of the type-I improvements in the Stern-Gerlach experiment. From Eqs. ($\ref{eq:b00}$),($\ref{eq:c22}$),($\ref{eq:d06}$), ($\ref{eq:d07}$), and ($\ref{eq:d08}$), we can obtain the standard deviations after postselection
\begin{equation}\label{e01}
\Delta p_z'=\sqrt{\frac{1}{4\Delta^2}+\frac{g^2(\sin^2{\theta}+\sin{\theta}\cos{\phi}e^{-2\Delta^2g^2})}{(1+\sin{\theta}\cos{\phi}e^{-2\Delta^2g^2})^2}},
\end{equation}
and
\begin{equation}\label{e02}
\Delta z'=\sqrt{\Delta^2-\frac{4g^2\Delta^4(\sin{\theta}\cos{\phi}+\sin^2{\theta}e^{-2\Delta^2g^2})e^{-2\Delta^2g^2}}{(1+\sin{\theta}\cos{\phi}e^{-2\Delta^2g^2})^2}}.
\end{equation}
Then we can get the type-I improvements of SNR
\begin{equation}\label{eq:b03}\begin{split}
I_{p_z,I}&=\frac{\Delta p_z \sqrt{P} |\delta p_z'|}{\Delta p_z'|g|}\\
&=\frac{|\cos{\theta}|\sqrt{K}}{\sqrt{2K^2+8\Delta^2g^2(\sin^2{\theta}+\sin{\theta}\cos{\phi}e^{-2\Delta^2g^2})}},
\end{split}\end{equation}
and
\begin{equation}\label{eq:b04}\begin{split}
I_{z,I}&=\frac{\Delta p_z \sqrt{P} |\delta z'|}{\Delta z'|g|}\\
&=\frac{|\sin{\theta}\sin{\phi}e^{-2\Delta^2g^2}|\sqrt{K}}{\sqrt{2K^2-8\Delta^2g^2(\sin{\theta}\cos{\phi}e^{-2\Delta^2g^2}+\sin^2{\theta}e^{-4\Delta^2g^2})}},
\end{split}\end{equation}
where $K=1+\sin{\theta}\cos{\phi}e^{-2\Delta^2g^2}$.
For $g=0.02\Delta p_z$, the values of $I_{p_z,I}$ and $I_{z,I}$ are pictured in Fig. 3. From Fig. 3, we can see that the maximum values of $I_{p_z,I}$ and $I_{z,I}$ are both approximately $1$. Although we can get very largely amplified shifts, the probabilities of obtaining the large shifts are very low. The low probabilities cancel the advantage of the amplification effect~\cite{brunner}. By numerical calculation, when $g \ll \frac{1}{2\Delta}$, we get that the maximum type-I improvements $I_{p_z,I}^{\max} = I_{z,I}^{\max} \approx 1.038$. In the Stern-Gerlach experiment, the post-selected state $\ket{\psi_f}$ is fixed. We have also searched for the values of the improvements for all PPS by numerical calculation, and obtained that the upper bound of them is also 1.038.  Those results mean that the SNR can not be effectively improved by the weak measurements.

However, this does not mean that the weak measurement methods are not helpful for actual experiments. As we know, measurement errors are divided into random errors and system errors~\cite{error1,error2}. The standard deviations considered here are the random errors which can be reduced by repeated measurements. For a given system error $\Delta_{\mathrm{sys}}$ which cannot be reduced by repeated measurements, the relative system error is $\delta_{\mathrm{sys}}=\frac{\Delta_{\mathrm{sys}}}{X}$~\cite{error1,error2}, where $X$ is the shift of the measuring device. By the idea of weak measurement, the shift $X$ could be largely amplified. So, the relative system errors could be effectively reduced by weak measurement in actual experiments.

\begin{figure}[t]
\centering \includegraphics[scale=0.25]{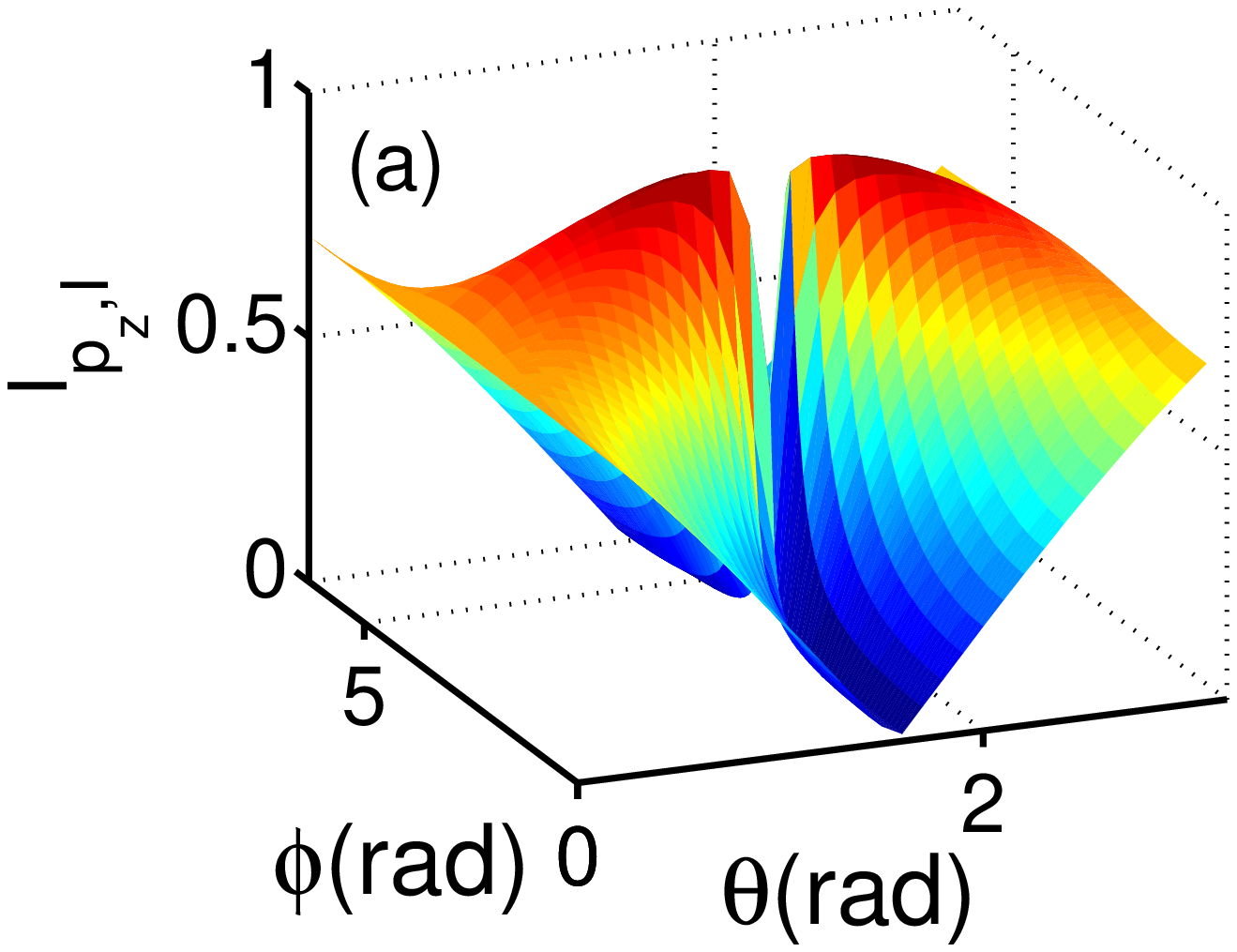} \includegraphics[scale=0.25]{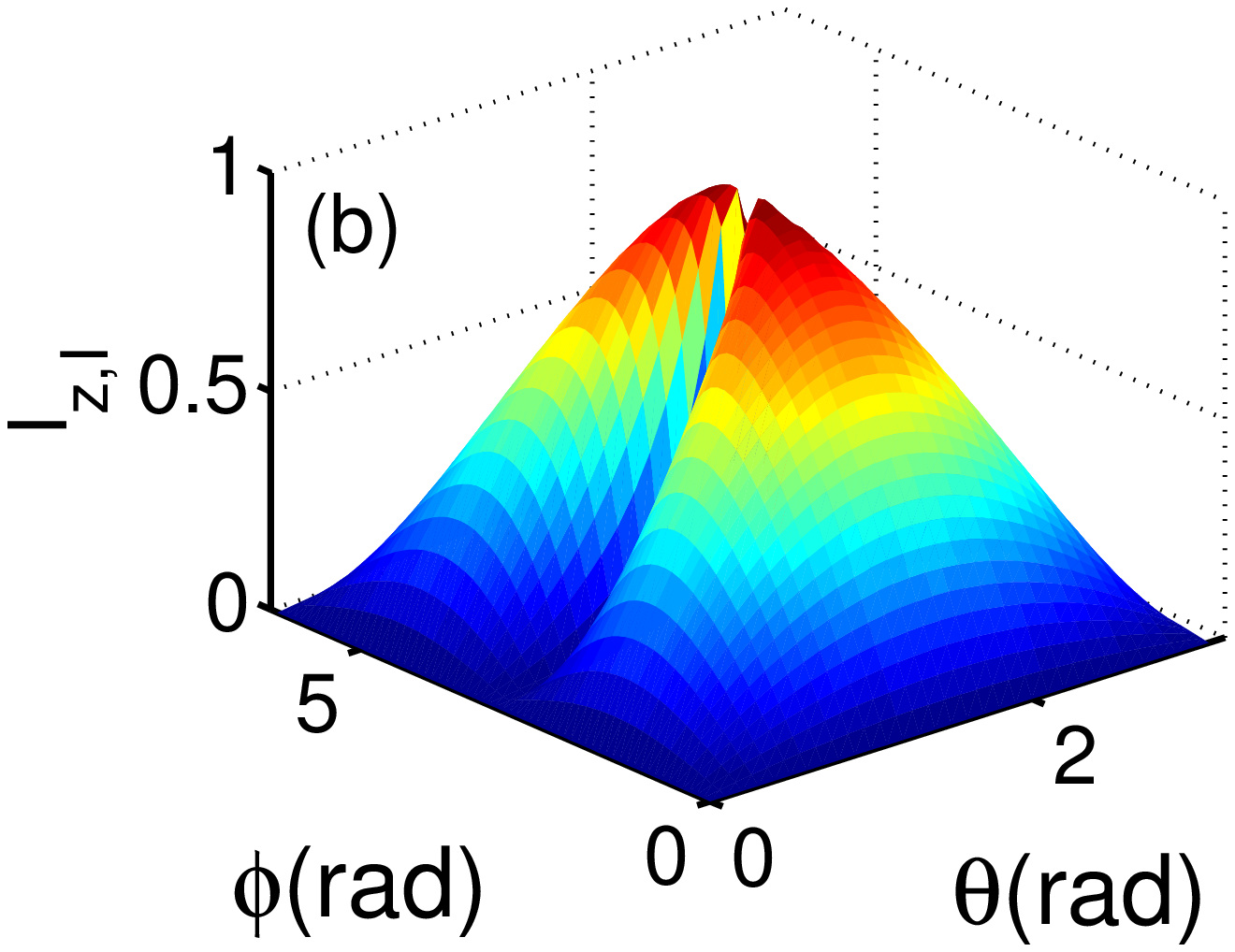}  \includegraphics[scale=0.25]{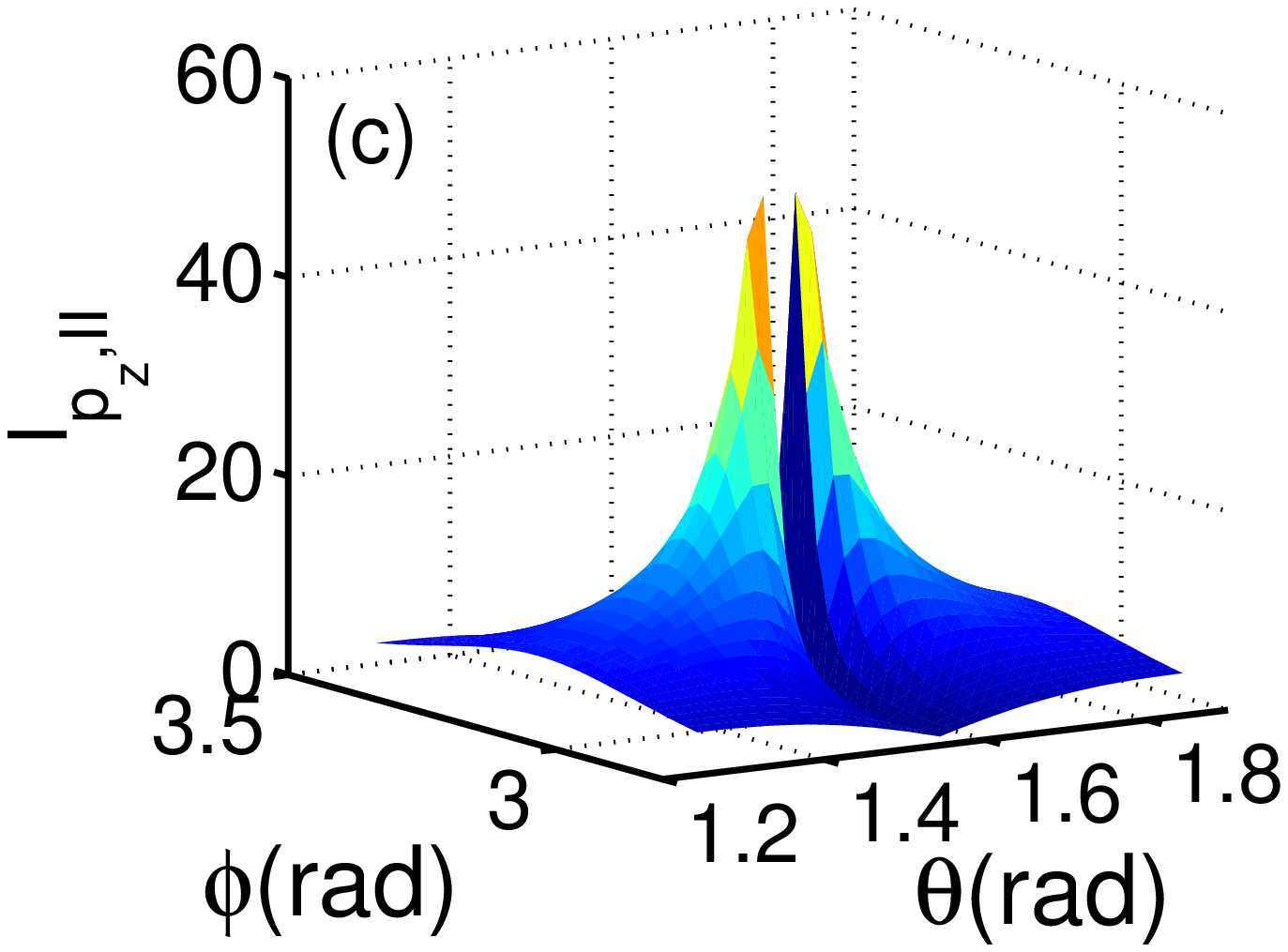} \includegraphics[scale=0.25]{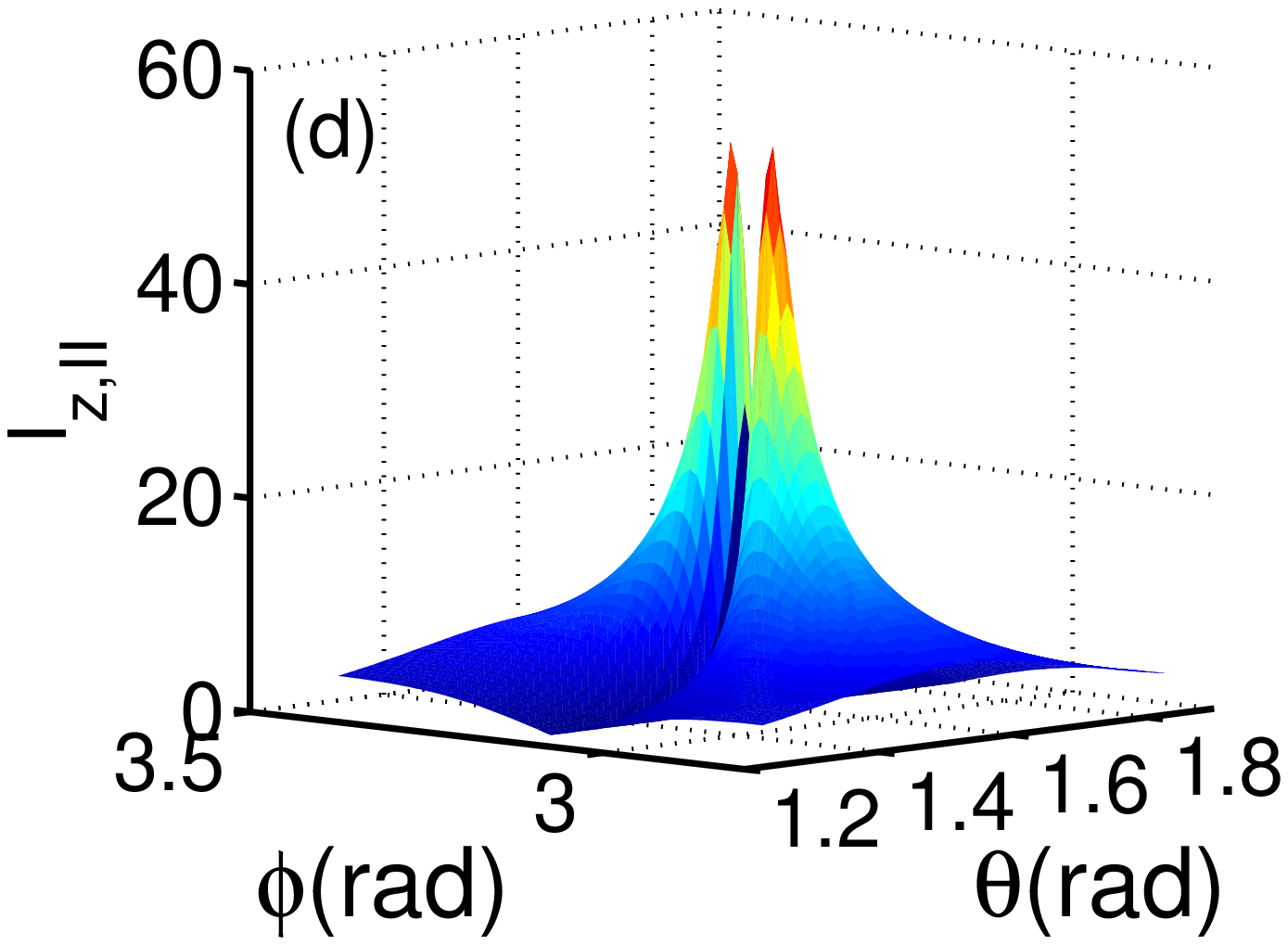} \caption{(Color online) The two types of improvements of SNR, when $g=0.02\Delta p_z$. From the four pictures, it can be seen that the values of $I_{p_z,I}$ and $I_{z,I}$ are both less than 1; meanwhile the maximum value of $I_{p_z,II}$ and $I_{z,II}$ are all approximatively equal to $53.7$.}
\label{fig:4}
\end{figure}

\subsection{Large improvements of the SNR when the probabilities reduced by postselection are not considered}
In the practical case, most experiments are limited by technical issues. Here we discuss the improvements of SNR under the detectors' saturation intensity technical limit~\cite{snr1}. In this circumstance, the input quantum systems are sufficient, and the numbers of the measurement repetition are determined by the detectors' saturation intensity. So, the numbers of repetition with and without postselection are almost the same. So the probability reduced by postselection needs not be considered in defining the improvements of SNR
\begin{equation}\label{eq:b01}\begin{split}
I_{p,II}=\frac{\Delta p |\delta p'|}{\Delta p'|a_{\max}g|},\\
I_{q,II}=\frac{\Delta p |\delta q'|}{\Delta q'|a_{\max}g|}.
\end{split}\end{equation}
They are referred to as type-II improvements of SNR, and in the definitions the probabilities reduced by postselection are not taken into account.
One can see that $I_{p,I}=\sqrt{P}I_{p,II}$ and $I_{q,I}=\sqrt{P}I_{q,II}$.

In the Stern-Gerlach experiment, the type II improvements of SNR are
\begin{equation}\label{eq:e03}
I_{p_z,II}=\frac{|\cos{\theta}|}{\sqrt{K^2+4\Delta^2g^2(\sin^2{\theta}+\sin{\theta}\cos{\phi}e^{-2\Delta^2g^2})}},
\end{equation}
and
\begin{equation}\label{eq:e04}
I_{z,II}=\frac{|\sin{\theta}\sin{\phi}e^{-2\Delta^2g^2}|}{\sqrt{K^2-4\Delta^2g^2(\sin{\theta}\cos{\phi}e^{-2\Delta^2g^2}+\sin^2{\theta}e^{-4\Delta^2g^2})}}.
\end{equation}

In Fig. 3, for $g=0.02 \Delta p_z$, it is shown that the maximum values  of $I_{p_z,II}$ and $I_{z,II}$ can approximately reach about 53.7 which means that the SNR can be efficiently improved.
When $g \ll \frac{1}{2\Delta}$, we can neglect the orders of $2\Delta^2g^2$ higher than $4\Delta^4g^4$ in Eq. (\ref{eq:e03}):
\begin{equation}\label{eq:a06}
I_{p_z,II}\approx \frac{|\cos{\theta}|}{\sqrt{(1+\sin{\theta}\cos{\phi})^2+F}}
\end{equation}
where $F=(4\Delta^2g^2{\sin^2{\phi}+8\Delta^4g^4\cos^2{\phi}})\sin^2{\theta}-4\Delta^2g^2\sin{\theta}\cos{\phi}$. By $\frac{\partial I_{p_z,II}}{\partial \phi}=0$, we get $\sin{\phi}=0$. As we want to search the maximum of $I_{p_z,II}$, by ensuring $\frac{\partial^2 I_{p_z,II}}{\partial \phi^2}<0$, we get $\cos{\phi}=-1$, then
\begin{equation}\label{eq:a07}
I_{p_z,II}=\frac{|\cos{\theta}|}{\sqrt{(1-\sin{\theta})^2+8\Delta^4g^4\sin^2{\theta}+4\Delta^4g^4\sin{\theta}}}.
\end{equation}
Denoting $L=I_{p_z,II}^2$, let $x=\sin{\theta}$, from $\frac{\partial L}{\partial x}=0$, we get $x=\sin{\theta}\approx 1-2\sqrt{3}\Delta^2g^2$. Substituting this solution into $L$, we obtain that $L_{\max} \approx \frac{\sqrt{3}}{6g^2\Delta^2}$, so $I_{p_z,II}^{\max} \approx \sqrt{\frac{\sqrt{3}}{6}}\frac{1}{|g\Delta|}$. For $g=0.02\Delta p_z$, we have $I_{p_z,II}^{max} \approx 53.7$ which is presented in Fig. 3. Through a similar procedure, we can also get $I_{z,II}^{max} \approx \sqrt{\frac{\sqrt{3}}{6}}\frac{1}{|g\Delta|}$. So, under the measurement repetition limited by the detectors' saturation intensity, the SNR can be greatly improved by weak measurement.

\section{Enhancement of the measurement sensitivity in weak measurements}
Using the idea of weak measurement, a series of tiny physical effects has been effectively detected in experiments~\cite{exp1,hall1,w3}. In those experiments, what has been actually measured could be regarded as the interaction strength between the systems and measuring devices. A natural question arises, whether weak measurement with preselection and postselection enhances the measurement precision in estimating the strength of a tiny interaction? In this section we will study the enhancements of the measurement sensitivity (MS) brought by weak measurement. We consider the enhancements with and without considering the probabilities reduced by postselection.

\subsection{Enhancement of the MS considering the probabilities reduced by postselection}

From the error propagation theory~\cite{error2,rev}, the error $\Delta g$ on the estimated coupling strength $g$ is
\begin{equation}\label{eq:d11}
 \Delta g \equiv \frac{\overline{\sigma_X}}{|\frac{\partial  X }{\partial g}|},
\end{equation}
where $X$ is the shift of the measuring device, and $\overline{\sigma_{X}}=\frac{\sigma_{X}}{\sqrt{N}}$ is the average standard deviation of $X$ after the measurement is repeated for $N$ times. If the interaction considered is $H=-g\delta(t-t_0)A\otimes q$, our purpose now is to estimate the value of the coupling strength $g$. Without postselection, the error $\Delta g$ is
\begin{equation}\label{eq:d12}
\Delta g = \frac {\Delta p}{|\frac{\partial \delta p}{\partial g}|\sqrt{N}}=\frac{\Delta p}{|\sum_{m}|\alpha_m|^2a_m|\sqrt{N}},
\end{equation}
where $\delta p$ is the shift of the pointer given in Eq. ($\ref{eq:d02}$) and $\Delta p$ is the standard deviation of momentum. By choosing an appropriate input state, we obtain the optimal error $\Delta g_{\mathrm{opt}}$ without postselection
\begin{equation}\label{eq:d13}
\Delta g_{\mathrm{opt}}=\frac{\Delta p}{|a_{\max}|\sqrt{N}},
\end{equation}
where $a_{\max}$ is the eigenvalue of observable $A$ with the maximum absolute value.
The errors $\Delta g_p$ and $\Delta g_q$ on the estimation of $g$ with postselection are
\begin{equation}\label{eq:d14}\begin{split}
\Delta g_p=\frac{\Delta p'}{|\frac {\partial \delta p'}{\partial g}|\sqrt{NP}}, \Delta g_q=\frac{\Delta q'}{|\frac{\partial \delta q'}{\partial g}|\sqrt{NP}},
\end{split}
\end{equation}
where $\delta p'$ and $\delta q'$ are given in Eqs. (\ref{eq:c8}) and (\ref{eq:c9}). With considering the probabilities reduced by postselection, the enhancements of MS are defined as
\begin{equation}\label{eq:d15}\begin{split}
E_{p,I}=\frac{\Delta g_{\mathrm{opt}}}{\Delta g_p}=\frac{\Delta p |\frac {\partial \delta p'}{\partial g}|\sqrt{P}}{|a_{\max}|\Delta p'},\\
E_{q,I}=\frac{\Delta g_{\mathrm{opt}}}{\Delta g_q}=\frac{\Delta p |\frac {\partial \delta q'}{\partial g}|\sqrt{P}}{|a_{\max}|\Delta q'}.
\end{split}\end{equation}
They are referred to as type-I enhancements of the MS, and in the definitions the probabilities reduced by postselection are taken into account.
If the value of the enhancement $E$ is bigger than 1, weak measurement could indeed improve the precision in estimating the interaction strength. Now, we consider the enhancements of MS in the Stern-Gerlach experiment. In this experiment, $a_{\max}=1$ and the type-I enhancements are
\begin{equation}\label{eq:e07}\begin{split}
E_{p_z,I} &=\frac{\Delta p_z|\frac{\partial \delta p'_z}{\partial g}|\sqrt{P}}{\Delta p_z'}\\
&=\frac{|\cos{\theta}||1+4\Delta^2g^2-4\Delta^2g^2/K|\sqrt{K}} {\sqrt{2K^2+8\Delta^2g^2(\sin^2{\theta}+\sin{\theta}\cos{\phi}e^{-2\Delta^2g^2})}},
\end{split}\end{equation}
and
\begin{equation}\label{eq:e08}\begin{split}
E_{z,I} &=\frac{\Delta p_z|\frac{\partial \delta z'}{\partial g}|\sqrt{P}}{\Delta z'}\\
&=\frac{|(1-4\Delta^2g^2/K)||\sin{\theta}\sin{\phi}e^{-2\Delta^2g^2}|\sqrt{K}}{\sqrt{2K^2-8\Delta^2g^2(\sin{\theta}\cos{\phi}e^{-2\Delta^2g^2}+\sin^2{\theta}e^{-4\Delta^2g^2})}},
\end{split}\end{equation}
where $K=1+\sin{\theta}\cos{\phi}e^{-2\Delta^2g^2}$. For $g=0.02 \Delta p_z$, the enhancements are pictured in Fig. 4. We can see that the values of $E_{p_z,I}$ and $E_{z,I}$ are both less than 1.

Comparing $E_{p_z,I}$ and $E_{z,I}$ with $I_{p_z,I}$ and $I_{z,I}$, we get that $E_{p_z,I}=|(1+4\Delta^2g^2-4\Delta^2g^2/K)|I_{p_z,I}$, and $E_{z,I}=|(1-4\Delta^2g^2/K)|I_{z,I}$.  For $g \ll \frac{1}{2\Delta}$, we have
$1-(1+4\Delta^2g^2-4\Delta^2g^2/K)^2\approx 8\Delta^2g^2/K(1-{2\Delta^2g^2}/{K}).$
Denoting $V={2\Delta^2g^2}/{K}$, by $\frac{\partial V}{\partial \phi}=0$ and $\frac{\partial V}{\partial \theta}=0$, we get the maximum values of $V_{\max} \approx 1$, and $0 \leq 1-{2\Delta^2g^2}/{K}$. As $K=1+\sin{\theta}\cos{\phi}e^{-2\Delta^2g^2} > 0$, we get $|(1+4\Delta^2g^2-4\Delta^2g^2/K)| \leq 1 $ and $ |(1-4\Delta^2g^2/K)| \leq 1 $. So, for $g \ll \frac{1}{2\Delta}$, we obtain that $E_{p_z,I} \leq I_{p_z,I}$, $E_{z,I}\leq I_{z,I}$.

We also search the values of the enhancements for all PPS by numerical calculation, and we find that the upper bound of the type-I enhancements is 1. Therefore, the MS can not be efficiently enhanced by weak measurement if the probability decrease due to postselection is taken into account. This was also pointed out in~\cite{hall1}.

 \begin{figure}[t]
 \centering \includegraphics[scale=0.25]{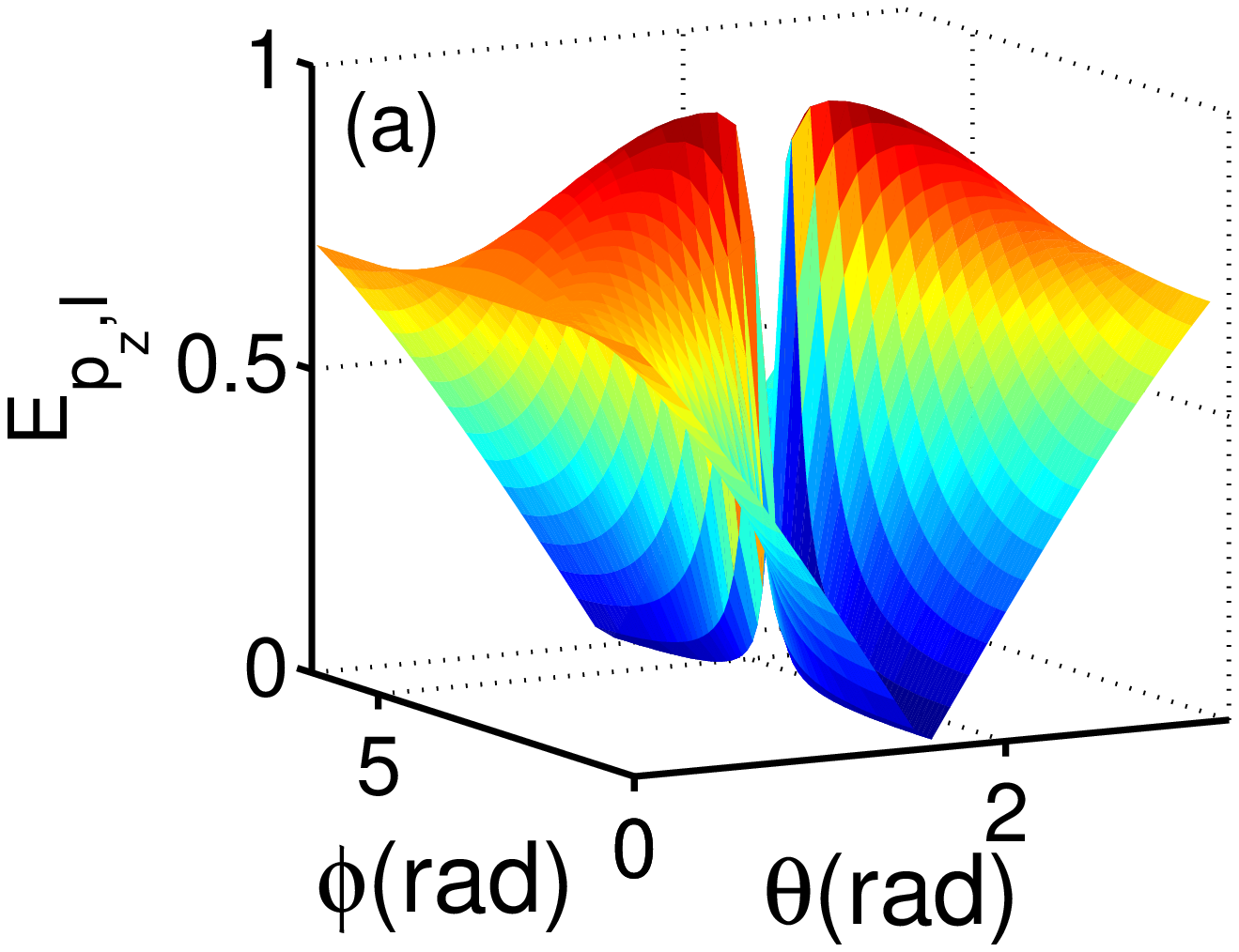} \includegraphics[scale=0.25]{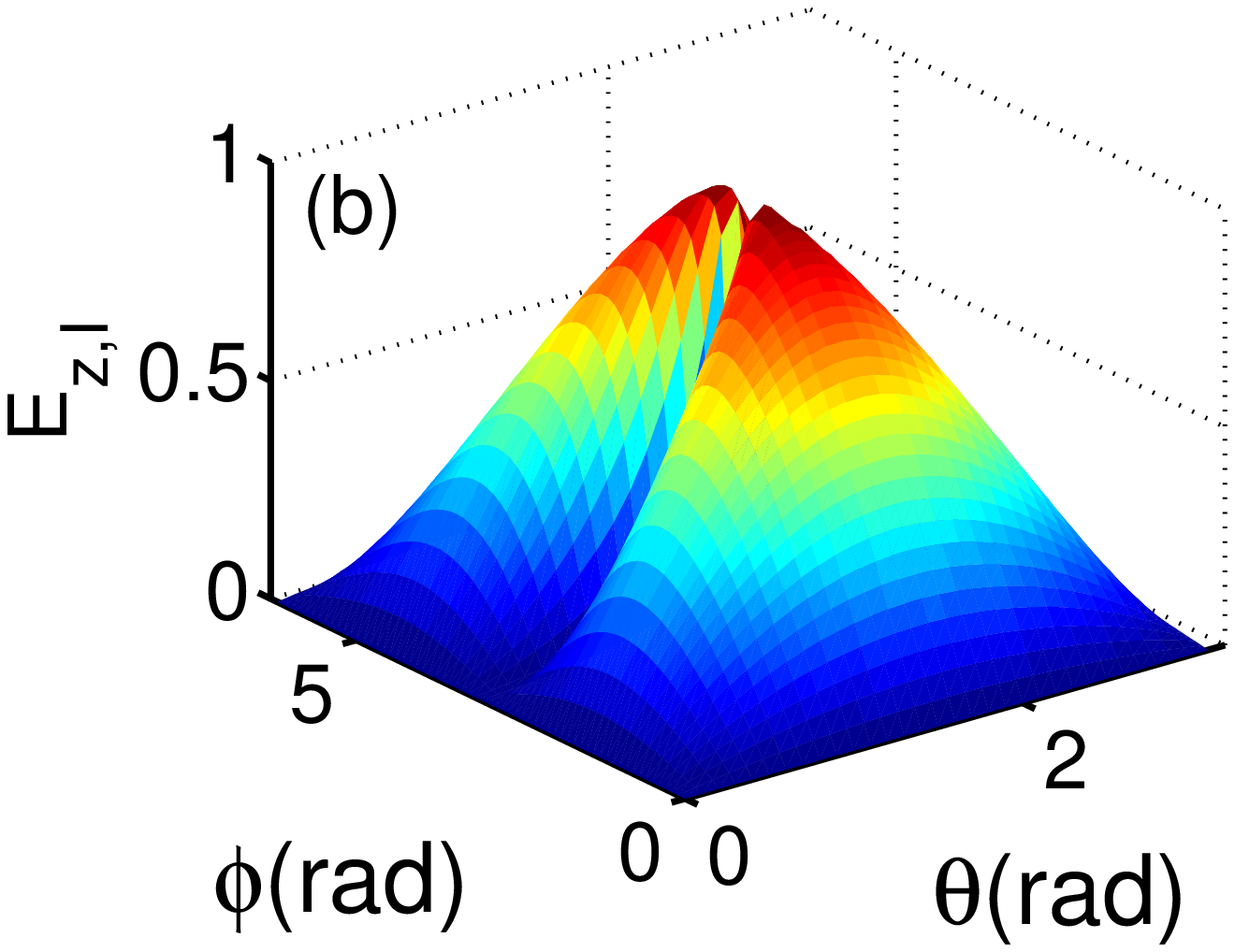} \includegraphics[scale=0.25]{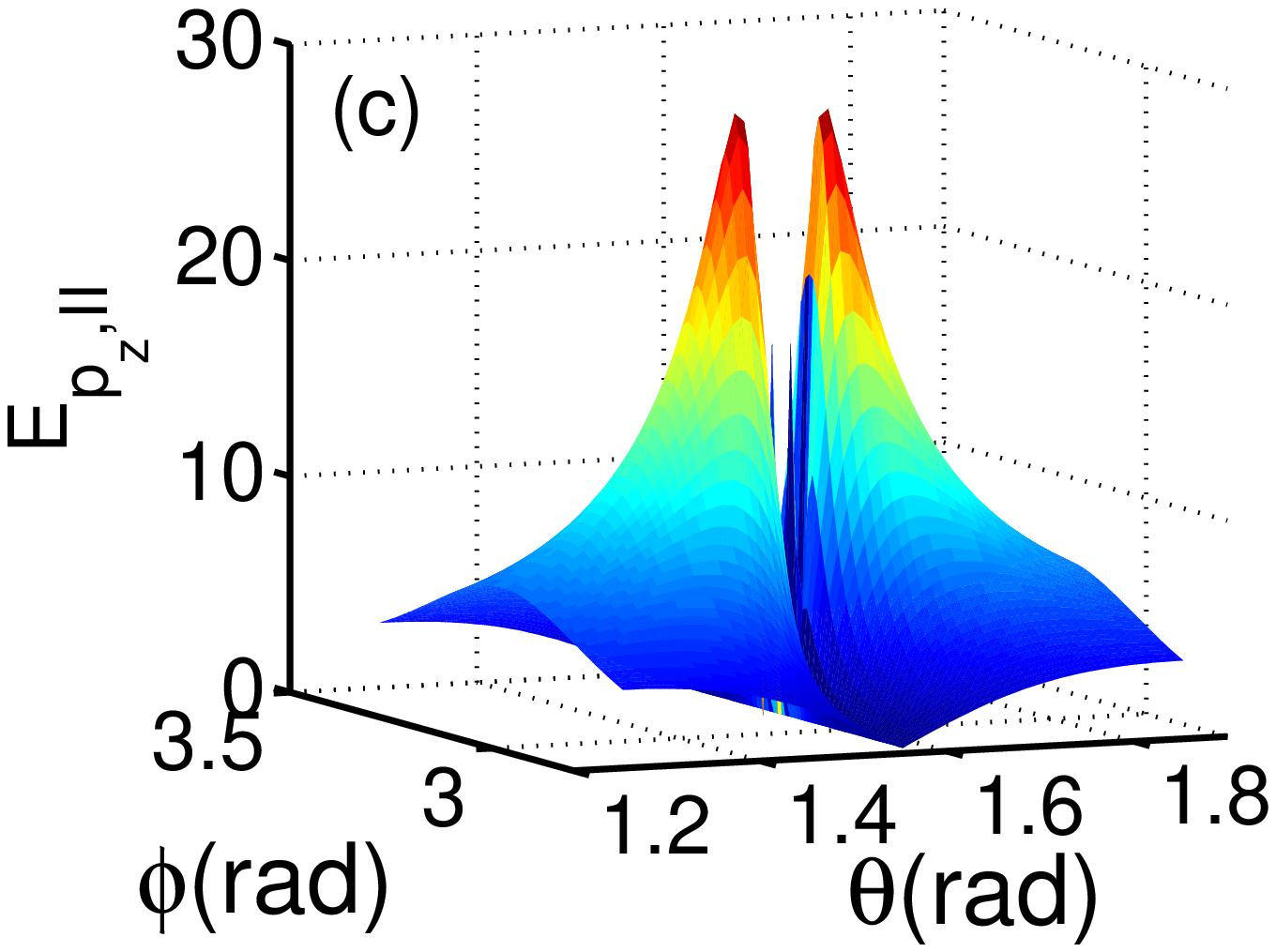} \includegraphics[scale=0.25]{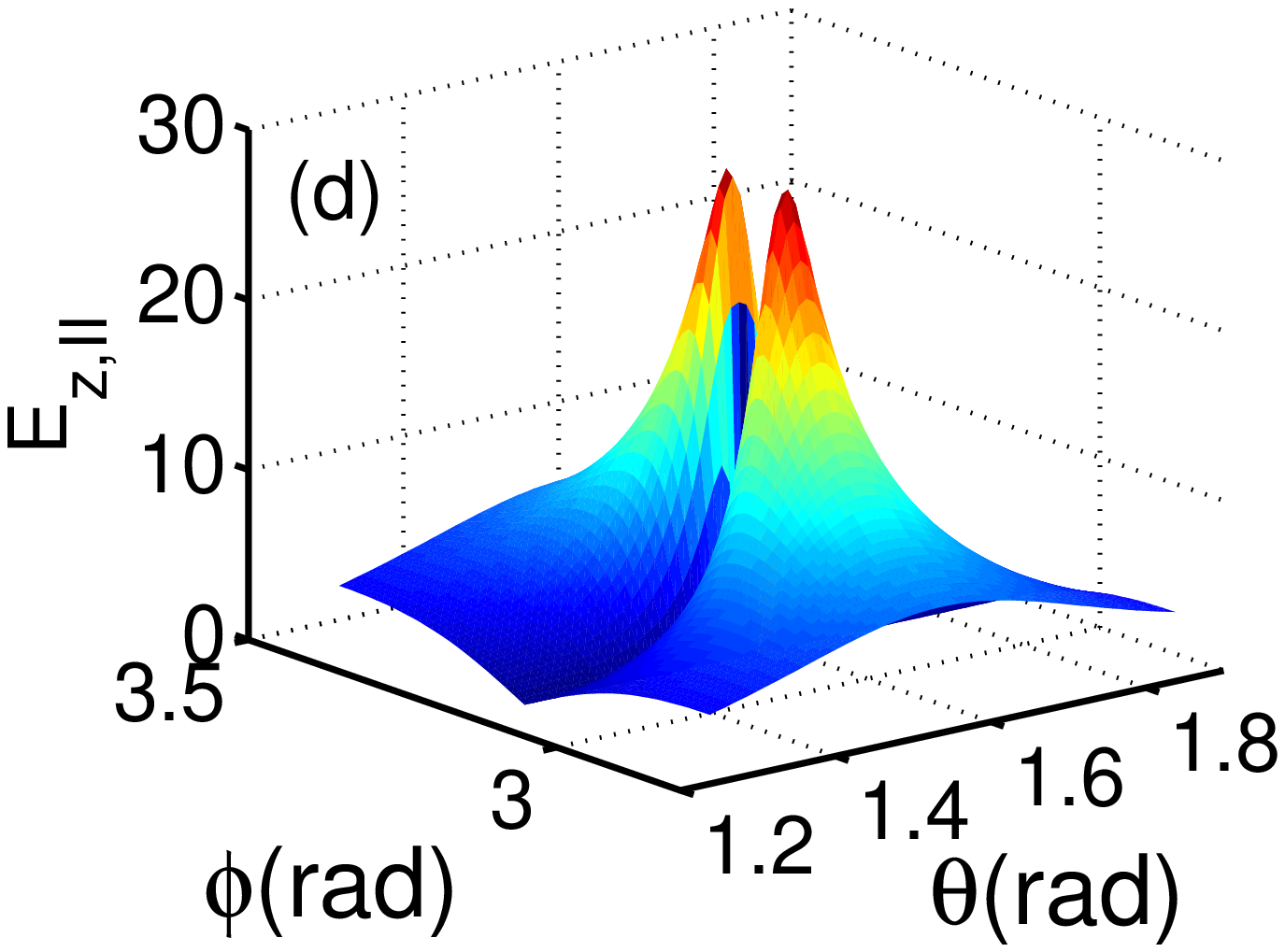}\caption{(Color online) The
enhancements of measurement sensitivity for  $g=0.02\Delta p_z$. For all PPS,  $E_{p_z,I}$ and $E_{z,I}$ are both less than 1; meanwhile the maximum values of $E_{p_z,II}$ and $E_{z,II}$ are both approximatively $28$.}
 \label{fig:3}
 \end{figure}

\subsection{Enhancement of the MS without considering the probabilities reduced by postselection}

In this section, we consider the same technical limit discussed in Sec. IV B. Under this technical limit, the number of the outcomes is determined by the saturation of the detectors without considering the probability decrease due to postselection, and we have an alternative definition of the enhancements of MS
\begin{equation}\label{eq:d16}\begin{split}
E_{p,II}=\frac{\Delta p |\frac {\partial \delta p'}{\partial g}|}{|a_{\max}|\Delta p'},\\
E_{q,II}=\frac{\Delta p |\frac {\partial \delta q'}{\partial g}|}{|a_{\max}|\Delta q'}.
\end{split}\end{equation}
They are referred to as type-II enhancements of MS, and in the definitions the probabilities reduced by postselection are not taken into account.
Similarly, we have that $E_{p,I}=\sqrt{P}E_{p,II}$ and $E_{q,I}=\sqrt{P}E_{q,II}$.

In the Stern-Gerlach experiment, we can have
\begin{equation}\label{eq:e09}\begin{split}
E_{p_z,II}=\frac{|\cos{\theta}||1+4\Delta^2g^2-4\Delta^2g^2/K|} {\sqrt{K^2+4\Delta^2g^2(\sin^2{\theta}+\sin{\theta}\cos{\phi}e^{-2\Delta^2g^2})}},
\end{split}\end{equation}
and
\begin{equation}\label{eq:e10}\begin{split}
E_{z,II}=\frac{|(1-4\Delta^2g^2/K)||\sin{\theta}\sin{\phi}e^{-2\Delta^2g^2}|}{\sqrt{K^2-4\Delta^2g^2(\sin{\theta}\cos{\phi}e^{-2\Delta^2g^2}+\sin^2{\theta}e^{-4\Delta^2g^2})}},
\end{split}\end{equation}
where $K=1+\sin{\theta}\cos{\phi}e^{-2\Delta^2g^2}$. We can see that $E_{p_z,II}=|(1+4\Delta^2g^2-4\Delta^2g^2/K)|I_{p_z,II}$, and $E_{z,II}=|(1-4\Delta^2g^2/K)|I_{z,II}$. As for $g \ll \frac{1}{2\Delta}$, we have $|(1-4\Delta^2g^2/K)| < |(1+4\Delta^2g^2-4\Delta^2g^2/K)| \leq 1$, so we can get $E_{p_z,II} \leq I_{p_z,II}$, $E_{z,II}\leq I_{z,II}$.
For $g=0.02\Delta p_z$, the values of enhancements $E_{p_z,II}$ and $E_{z,II}$ are pictured in Fig. 4. It can be seen that the maximum value of the enhancements is about 28, which means the MS can be largely improved by the weak measurements when the probability decrease due to postselection is not considered. For $g \ll \frac{1}{2\Delta}$, by numerical calculation, we get $E_{p_z,II}^{\max} \approx E_{z,II}^{\max} \approx \frac{0.28}{g\Delta}$, which implies that weak measurements can effectively enhance MS when the probability decrease due to postselection is not considered, such as for practical cases when the measurement repetition frequency is determined by the intensity saturation of the detectors. Significantly enhanced MS has been obtained in observing the spin hall effect of light~\cite{hall1,hall2}.

\section{conclusions}
We have studied quantum measurement with preselection and postselection, and have given the precise expressions of the measurement results without any restriction on the system-device interaction strength. For the very weak interaction, a significant amplification effect can be obtained. For a qubit system, the maximal pointer shifts of momentum and position are obtained. When the interaction between the system and device is strong, we have also obtained the ideal quantum measurement results.

We also have studied the improvements of the SNR and the enhancements of the MS by weak measurements. We have shown that weak measurements cannot effectively improve the SNR and the MS when the probability decrease due to postselection need to be considered; while for practical cases when the probability reduced by postselection need not be considered, weak measurements can significantly improve both the SNR and the MS. In addition, the large shifts obtained in weak measurements can also effectively reduce the relative system error. The idea of weak measurement is very useful in practical experiments.

\section*{Acknowledgments}
This work is financially supported by the National Natural Science Foundation of China(Grants No. 11075148, and No. 11175063), the Fundamental Research Funds for the Central Universities, CAS, and the National Fundamental Research Program, China.

\end{document}